\documentclass[aip,reprint]{revtex4-1}

\pdfoutput=1

\usepackage{graphicx}
\usepackage{amsmath}
\usepackage{url}
\newcommand{\degrees}{$^\circ$}
\newcommand{\micron}{$\mu$m\ }
\newcommand{\Ohm}{$\Omega$\ }

\begin{document}
\title{Non-linear radial spinwave modes in thin magnetic disks}
\author{M. Helsen}
\email{Mathias.Helsen@UGent.be}
\affiliation{Department Solid State Sciences, Ghent University, Krijgslaan 281/S1, 9000 Ghent, Belgium}
\author{A. Gangwar}
\affiliation{Department of Physics, Universit\"{a}t Regensburg, Universit\"atsstrasse 31, 93040 Regensburg, Germany}
\author{J. De Clercq}
\affiliation{Department Solid State Sciences, Ghent University, Krijgslaan 281/S1, 9000 Ghent, Belgium}
\author{A. Vansteenkiste} 
\affiliation{Department Solid State Sciences, Ghent University, Krijgslaan 281/S1, 9000 Ghent, Belgium}
\author{B. Van Waeyenberge}
\affiliation{Department Solid State Sciences, Ghent University, Krijgslaan 281/S1, 9000 Ghent, Belgium}


\date{\today}
\begin{abstract}
We present an experimental investigation of radial spin-wave modes in magnetic nano-disks with a vortex ground state. The spin-wave amplitude was measured using a frequency-resolved magneto optical network analyzer, allowing for high-resolution resonance curves to be recorded. It was found that with increasing excitation amplitude up to about 10\,mT, the lowest-order mode behaves strongly non-linearly as the mode frequency redshifts and the resonance peak strongly deforms. This behavior was quantitatively reproduced by micromagnetic simulations. % which also reveal that the non-linearity is 
%caused by a transformation of the linear spin-wave into a self-focusing soliton. 
At higher excitation the spinwaves are transformed into a soliton by self-focusing, and  collapse onto the vortex core, dispersing the energy in short-wavelength spinwaves. Additionally, this process can lead to switching of the vortex polarization through the injection of a Bloch point.
\end{abstract}
\maketitle
%\section{Introduction}
The study of the static and dynamic properties of micron and sub-micron sized magnetic platelets is not only necessary for the development of new technological applications, but is also interesting for fundamental reasons, as they can be simple model systems to investigate magnetic interactions and the complex dynamics involved. This is especially true for platelets with a magnetic vortex ground state configuration. It is the simplest, non-trivial configuration in a magnet and a building block of more complex states. Such vortex configurations know several modes of excitation. The lowest frequency mode is the gyrotropic mode (typically in the sub-GHz range)\cite{guslienko:2006,choe:2004}, and corresponds to cyclic motion of the vortex around the center of the structure.  This mode has already been studied extensively, especially because its excitation can lead to switching of the vortex core\cite{bvw:2006,vansteenkiste:2008,kamionka:2010}. At higher frequency (typically several GHz), azimuthal modes exist\cite{park:2005,neudecker:2006}. These modes hardly move the vortex core, but the magnetization is excited out-of-plane with a periodicity in function of the azimuthal angle and the wave vector $\vec{k}$ parallel to the magnetization $\vec{m}$.  Because of azimuthal symmetry, this mode is divided in clockwise (CW) and counter clockwise (CCW) modes. However, the coupling with the vortex core, lifts the degeneration between these modes and can result in uni-directional switching of the vortex core\cite{kammerer:2011}.
The third excitation mode is the radial spin wave mode. Here the out-of-plane excursion amplitude of the magnetization is now a function of the radius. As the wave vector is perpendicular to the magnetization direction, this is a Damon-Eschbach mode. Until now, experimental work has only focused on the low excitation regime where the response is linear \cite{buess:2004,awad:2010,vogt:2011,castel:2012}, but recent numerical studies predict that at higher excitation levels vortex core switching \cite{yoo:2012,Pylpovskyi:2013} and non-linear phenomena can appear \citep{Pylpovskyi:2013,Moon:2014}. Here, we present an experimental investigation of the non-linear regime of the radial spinwave mode and further elaborate on the non-linear phenomena appearing near the core switching threshold.

\begin{figure}
    \includegraphics[width=5.5cm]{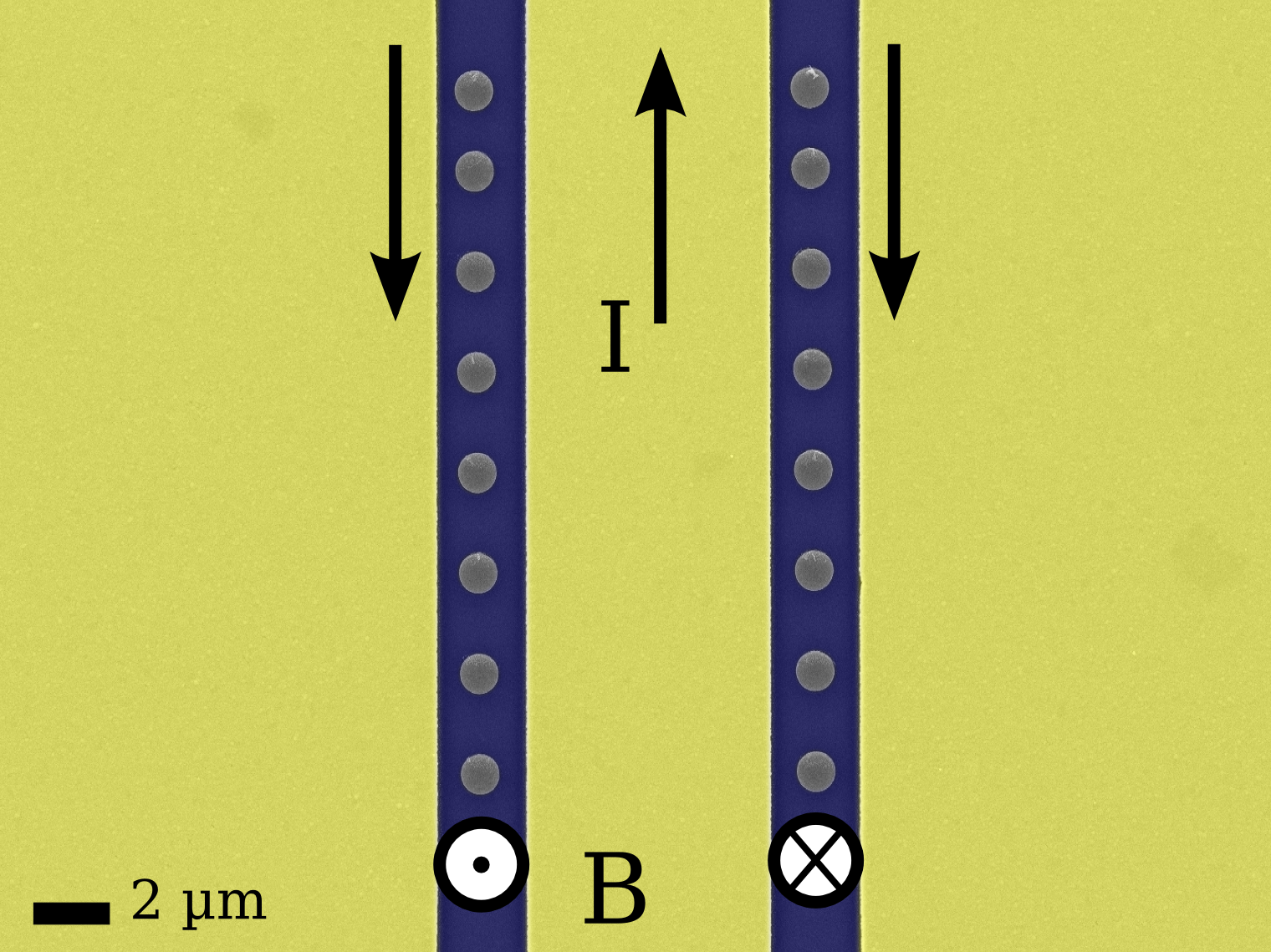}
    \caption{False-color scanning electron micrograph of the investigated sample. 1\,\micron wide, 10\,nm thick Permalloy disks are placed in the gaps of a coplanar waveguide used to generate an out-of-plane oscillating magnetic field in the GHz range.}
    \label{semimage}
\end{figure}

%\section{Experimental}
Our samples consisted of 1\micron Permalloy discs, 10nm in thickness located in the gap of a gold coplanar waveguide (CPW). The thickness was kept thin so the onset of non-linear behavior was at relatively low excitation strengths, as was found by micromagnetic simulations. The disks were separated by 1\micron to avoid dipolar coupling.
The nanostructures were created directly on a silicon substrate using thermal evaporation and electron beam lithography. The entire sample was covered with a conformal layer of aluminum oxide (7nm) by Atomic Layer Deposition (ALD) to prevent oxidation. Samples were inspected using Atomic Force Microscopy and Scanning Electron Microscopy (Fig. \ref{semimage}) to determine the lateral size of the discs. 

To study the magnetization dynamics we have used a frequency resolved magneto-optical Kerr microscope, described in\cite{helsen:2014}, which measures the RMS out-of-plane magnetization variation in the frequency domain. The method allows an arbitrary frequency resolution and the ability to resolve single nano-elements. To this end a probing laser beam, with a beam waist of about 500nm, is focused on the investigated nano disk. This allows to directly measure resonance curves of individual elements, in contrast to e.g. VNA-FMR where the integrated signal of several discs may lead to artificial line broadening.

A continuous wave RF current from a signal generator flowing through the center conductor and ground plane of the CPW generates a magnetic field that is perpendicular to the substrate in the center of the gap, but that acquires an in-plane component at the edge of the gap. The dimensions of the CPW were tuned to reach a compromise between field homogeneity (lower bound on the size of the gap) and magnetic field strength for a given RF current (upper bound). A too small gap would yield excessive in-plane excitation, possibly exciting azimuthal spin waves. Though the setup can not measure these modes (the spatially averaged out-of-plane excitation is zero) they could couple to the radial modes, distorting the spectra. Micromagnetic simulations were performed to confirm that for the dimensions chosen no azimuthal spin waves would be excited, i.e. azimuthal modes were separated in frequency from the radial modes and the effect of the small in-plane field was negligible. The width of the center conductor of the CPW was computed to produce a 50\Ohm transmission line. The current distribution through the CPW was calculated and when combined with the measurements of the power transmitted through the CPW, yields the local RF magnetic field.

\begin{figure}
    \includegraphics[width=8.5cm]{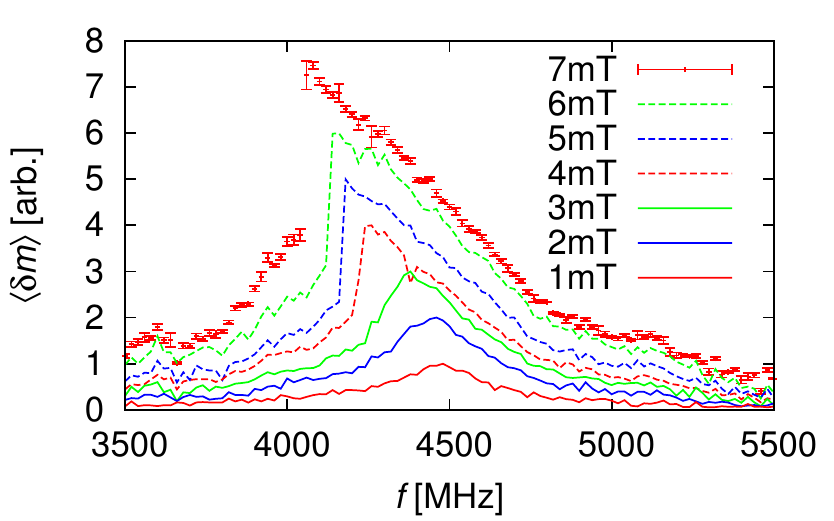}
    \caption{Spatially averaged out-of-plane magnetization for different out-of-plane excitation strengths and frequencies. At low field strengths, the resonance curve agrees well with a Lorentzian. When the excitation is increased, the resonance shows redshifting and acquires a skewness at the highest excitation fields. Both phenomena are indicative of non-linear effects. For the data at the highest excitation, standard deviations derived from repeated measurements are included.}
    \label{transition}
\end{figure}

The measured resonance curves at different excitation strengths between 1 and 7\,mT are shown in Fig. \ref{transition}. The resonance frequency of 4448.5\,MHz $\pm$ 1.3\,MHz at the lowest excitation level, in the linear regime, agrees well with the predicted value of 4500\,MHz for this aspect ratio ($\beta = t/R = 0.02$)\cite{awad:2010,vogt:2011}.

At high excitation strengths the resonance curves are systematically shifted towards lower frequencies. Similar redshifting has previously been observed for the vortex gyration mode\cite{buchanan:2007,drews:2012}. Additionally, a strong skewing of the spectrum at higher excitation is observed. However, as we do not expect the vortex core to be displaced for pure radially symmetric excitation, it can not be related an-harmonicity of the dipolar vortex potential \cite{drews:2012,Sukhostavets:2013}, and another non-linear phenomenon must be responsible for the deformation of the resonance curve. The frequency sweeps were performed in both directions, but no hysteresis (that could indicate a Duffing-oscillator type of behavior \citep{Moon:2014}) was found.

%\section{Discussion}
To gain more insight in the behaviour of the system at high excitation levels, micromagnetic simulations were used. The in-house developed micromagnetic code MuMax3\cite{mumax} was used to perform 2D simulations with a cell size of 4$\times$4$\times$10\,nm$^3$ and a simulation box of 256$\times$256$\times$1 cells. The material parameters were tuned to match the measured resonance data in the linear
regime (magnetic field amplitude of 1\,mT): saturation magnetization $M_\text{s} =$ 690\,kA/m, exchange stiffness $A_\text{ex} = 13\cdot10^{-12}$\,J/m, no magnetocrystalline anisotropy and damping parameter $\alpha$=0.008. For each simulation, the excitation field was ramped up linearly from zero to the nominal value, after which it was kept constant for 50 periods.
During this time the spatial average of the out-of-plane magnetization was saved. In post processing, the spin wave amplitude was calculated by multiplying this average with a sine and cosine at the excitation frequency and deriving the magnitude. This equivalent to the quantity that what was measured experimentally. The frequency increment was taken the same as in the experiments, 20\,MHz.

\begin{figure}
    \includegraphics[width=8.5cm]{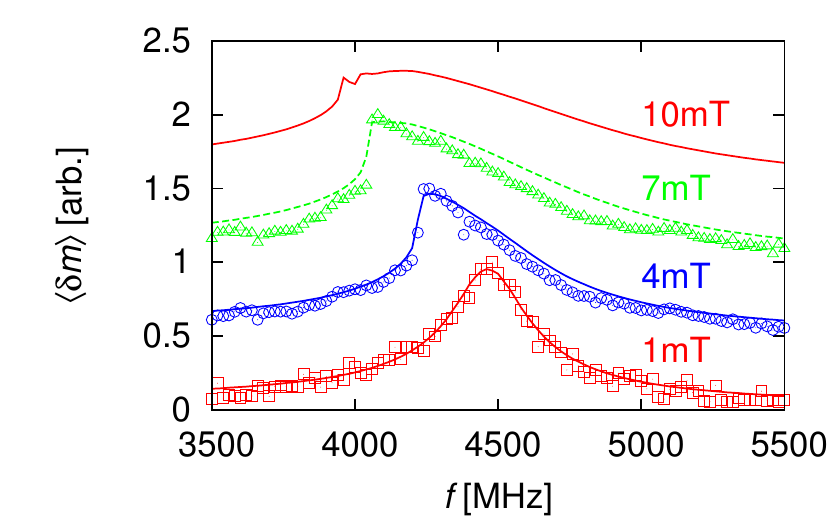}
    \caption{Comparison between the experimental data (symbols) and simulations (lines). At the bottom, the resonance in the linear regime (1\,mT peak excitation amplitude) is shown, where the material parameters were fitted ($M_\text{S} =$ 690kA/m, $\alpha = 0.008$). The other experimental data at 4\,mT excitation (circles) and 7\,mT (triangles) agree well with the simulations without further fitting. The highest spectrum at 10\,mT excitation shows a reduction of the response at resonance. All curves are scaled to unity and offset for clarity.}
    \label{sims_vs_exp}
\end{figure}

The results of the simulations and the comparing experimental data are shown in Fig. \ref{sims_vs_exp}. At low excitation levels the curve fits a Lorentzian and there is no indication of non-linear behavior. In this linear regime the radial spinwave is a pure standing wave.
% TODO: Y-axis is confusing

When the excitation is increased, the simulated resonance peak also shifts towards lower frequencies and acquires a strong skew (Fig. \ref{sims_vs_exp}). The good agreement with the experiment gives confidence that the simulations can provide more insight. So we compare the spatial mode profiles at different excitation levels. Up to 7\,mT, no significant deviation from the linear-regime profile (low excitation) is observed. However, at that point, the magnetization at the antinode already acquires precession angle of 10\degrees . Such wide-angle precession falls outside the regime covered by linear spin wave theory where only first order terms in the dynamic magnetization are considered, explaining the observed non-linearities.

\begin{figure}
    \includegraphics[width=8.5cm]{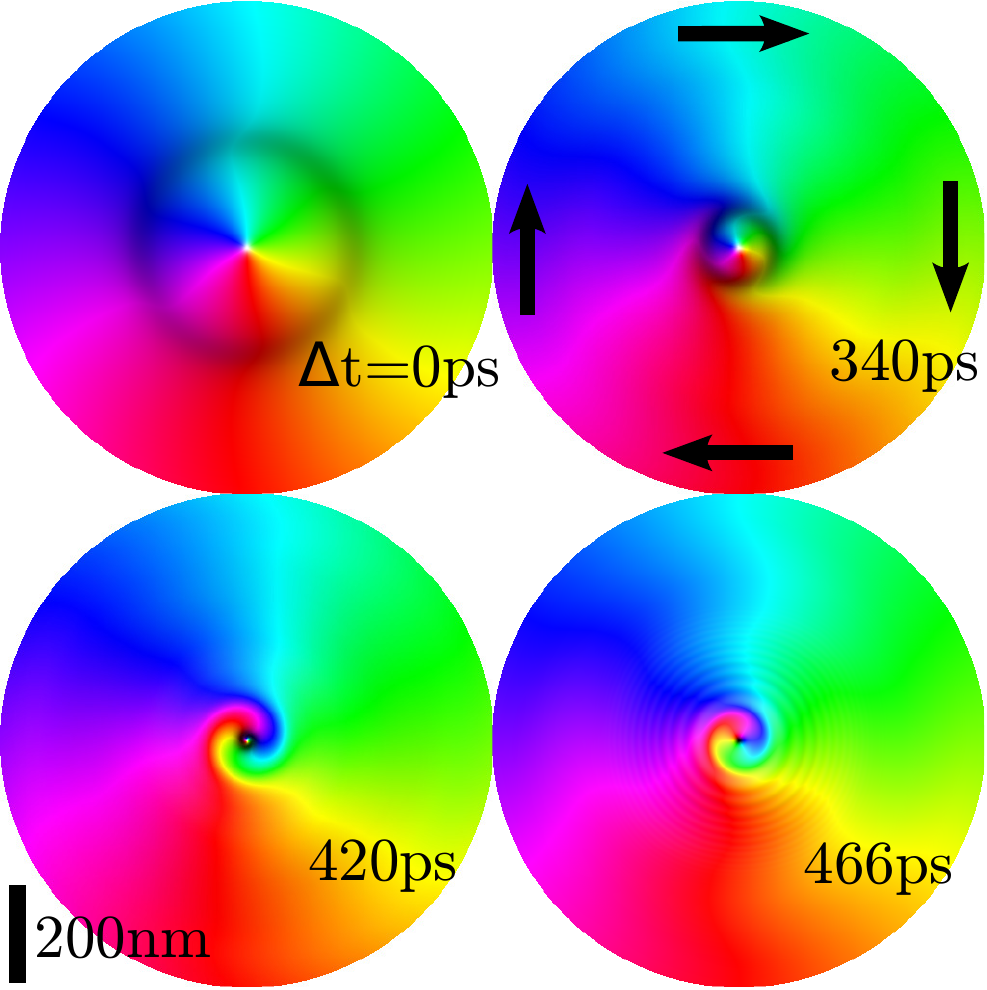}
    \caption{Snapshots of the magnetization at 10\,mT excitation, illustrating the formation of a soliton
	and subsequent vortex core switching. After core switching, the soliton breaks up into 
	short-wavelength spinwaves (last panel).
	The in-plane magnetization angle is color coded (see arrows in the second panel). White and black indicate
	the magnetization points up or down, respectively.
	}
    \label{soliton}
\end{figure}

Only when the excitation is increased to 10\,mT, just beyond the maximum amplitude that could be reached in the experiment, an additional feature in the spectrum appears: a depression near the maximum. From the temporal evolution of the magnetization (Fig. \ref{soliton}), one can see that the radial spinwave shows self-focusing, creating strongly localized soliton. This process has also been observed in a non-confined geometries\cite{bauer:1998}. The strong out-of-plane magnetization of the soliton forces the in-plane magnetization in its vicinity to precess, with a local inversion of the in-plane magnetization as a result. When the soliton reaches the center of the structure, it collapses and the energy is released by the emission of shorter wavelength spin waves. This can explain the depression in the spectrum, as we average out the response over the entire surface and hence do not detect those spinwaves. The simulations also show that this collapse switches the vortex core polarization, in agreement with \cite{yoo:2012,wang:2012}.  
The transformation of the soliton into shorter wavelengths is not dependent on core switching, however. Simulations have shown that when the core is artificially pinned (with a 20T field at the center of the core) or if a small hole is introduced in the center, the soliton still breaks up into short wavelength spinwaves.

%In comparison to previous work\cite{yoo:2012,wang:2012} we find that switching
%starts at a much lower threshold than has been estimated so far. 

\begin{figure}
    \includegraphics[width=8.5cm]{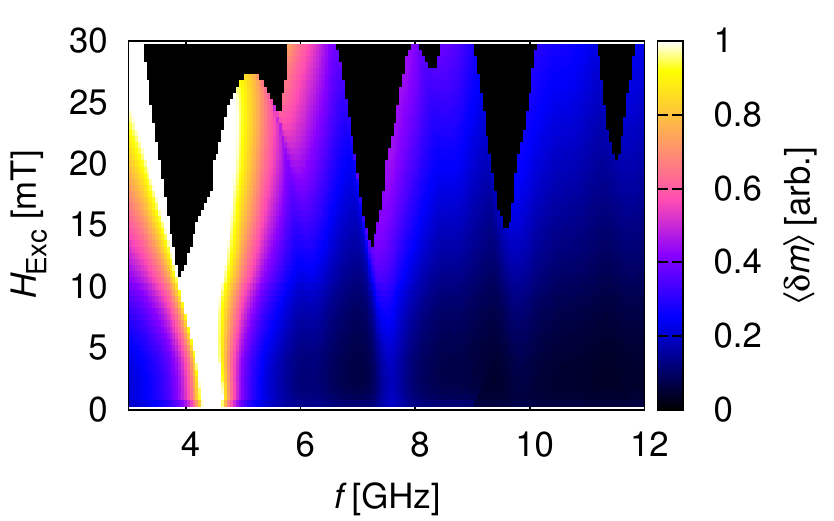}
    \caption{
	The switching diagram for a 1\micron diameter, 10nm thick permalloy dot. Areas that are colored black show vortex core switching.
	Higher order modes are also visible and can also lead to switching.
    }
    \label{switching_heatmap}
\end{figure}

A vortex core switching diagram was also calculated. To this end simulations were performed for various excitation strengths and frequencies and the core polarization was recorded. If the polarization had inverted within 25 excitation periods, this point in parameter space was marked as switching. The result is shown in Fig. \ref{switching_heatmap}. From the switching diagram it is clear that the higher order modes require more excitation to switch the vortex core. The excitation is uniform in space, thus the lowest order, uniform mode shows the greatest overlap, with successive modes showing less effective overlap with the excitation. 

The previous numerical studies \cite{yoo:2012,wang:2012} have raised some doubts on nature of the switching mechanism, as no clear Bloch-point was observed. However, the change in the topology charge during a core reversal always requires a Bloch-point. In the case of a radial asymmetric excitation, a short-lived vortex-antivortex pair is created\citep{bvw:2006} and the Bloch-point only appears only during the annihilation process.  For a radial symmetric excitation, the switching proceed should proceed in a punch through fashion with the injection of a Bloch-point\cite{thiaville:2003}. For this purpose we also investigated the switching process in the lowest mode through 3D simulations using 512$\times$512$\times$4 cells. It was found that the switching process is of a punch-through type where a Bloch-point is injected and inverts the core polarization, similar to\cite{thiaville:2003}. The actual time necessary for switching was found to be 1\,ps.

%\section{Conclusion}
In conclusion, we investigated radial spinwaves in a thin magnetic disc with a vortex ground state. Experimentally we have found both redshifting and skewing of the spectrum at high excitation amplitudes. 
Both simulations and measurements have shown that the lowest order radial spinwave mode can be driven into a non-linear regime. When the excitation is high enough, the confined Damon-Eschbach spinwaves can self-focus into a soliton.  When it then approaches the center of the disc, where it can switches the vortex core polarization and collapses with the emission of short wavelength spinwaves. This is characterized by a strongly skewed and redshifted spectrum with a reduced response above the switching threshold.

\bibliography{biblio}

\end{document}